# Fabrication of polarization-independent single-mode waveguides in lithium niobate crystal with femtosecond laser pulses

Jia Qi[1,2], Peng Wang[1,2], Yang Liao[1,4], Wei Chu[1], Zhengming Liu[1,2], Zhaohui Wang[1], Lingling Qiao[1], Ya Cheng[1,3,5]

[1]*State Key Laboratory of High Field Laser Physics, Shanghai Institute of Optics and Fine Mechanics, Chinese Academy of Sciences, Shanghai 201800, China*
[2]*School of Physical Science and Technology, Shanghai Tech University, Shanghai 200031, China*
[3]*State Key Laboratory of Precision Spectroscopy, East Normal University, Shanghai 200062, China*
[4]*superliao@vip.sina.com*
[5]*ya.cheng@siom.ac.cn*

**Abstract:** We report on fabrication of depressed cladding optical waveguides buried in lithium niobate crystal with shaped femtosecond laser pulses. Depressed cladding waveguides of variable mode-field sizes are fabricated by forming the four sides of the cladding using a slit-beam shaping technique. We show that the waveguides fabricated by our technique allows single-mode propagation of the light polarized in both vertical and horizontal directions.

## 1. Introduction

Femtosecond laser micromachining has enabled in-volume materials processing that opens horizons for innovative photonic and microfluidic applications [1-3]. In particular, fabricating optical waveguides with femtosecond laser direct writing uniquely allows formation of three-dimensional (3D) photonic circuits in various transparent materials (i.e., glass, crystals and polymers) in a continuous single-step processing manner [4-9]. Typically, waveguides written in fused silica and borosilicate glass consist of cores with increased refractive index that overlap the center of focal spots of the writing beams. However, irradiation of femtosecond laser pulses in lithium niobate can induce a negative refractive index change [9].

Owing to its large nonlinear optical and electro-optic coefficients, lithium niobate has been regarded as a platform for some classical and quantum photonic applications [10]. Various waveguide geometries have been demonstrated in this substrate, such as waveguides with positive refractive-index change (type I) [11,12], stress-based double-line waveguides (type II) [13,14], and depressed cladding waveguides [15,16]. Due to the capabilities of forming symmetrical cross sections that can support both polarization modes and preserving the nonlinear optical properties of the bulk crystals, the depressed cladding waveguides are now under intensive investigations. To form the depressed cladding waveguides, one can produce a cladding region of reduced refractive index through the irradiation of femtosecond laser pulses. For example, in the transverse writing scheme, the claddings can be formed by parallelly overlapping a large number of tracks written by the focused femtosecond laser pulses along the boundaries of the waveguiding areas [6,15-17]. The large thicknesses of the claddings formed in this way leads to the difficulties in reducing the size of the mode-field. Furthermore, constructing evanescent-coupling-based photonic circuits such as beam splitters and Mach-Zender interferometers would be difficult as a result of the thick claddings. At last, since the fabrication resolution in the axial direction critically depends on the numerical aperture (NA) of the objective lens, it is often required to use high NA objectives to produce the depressed claddings. The high NA objectives intrinsically limit the fabrication depth due to their limited working distances.

Here, we propose an approach that can overcome the above-mentioned challenges. We show that by transversely writing the depressed cladding waveguides consisting of only four tracks formed with slit-shaped femtosecond laser beams in lithium niobate substrates, single-mode light guiding can be achieved for both TE and TM modes at a wavelength of 1550 nm. Moreover, the waveguides allow for reducing the mode field dimensions down to that compatible with direct coupling to the standard single-mode optical fibers.

## 2. Experimental

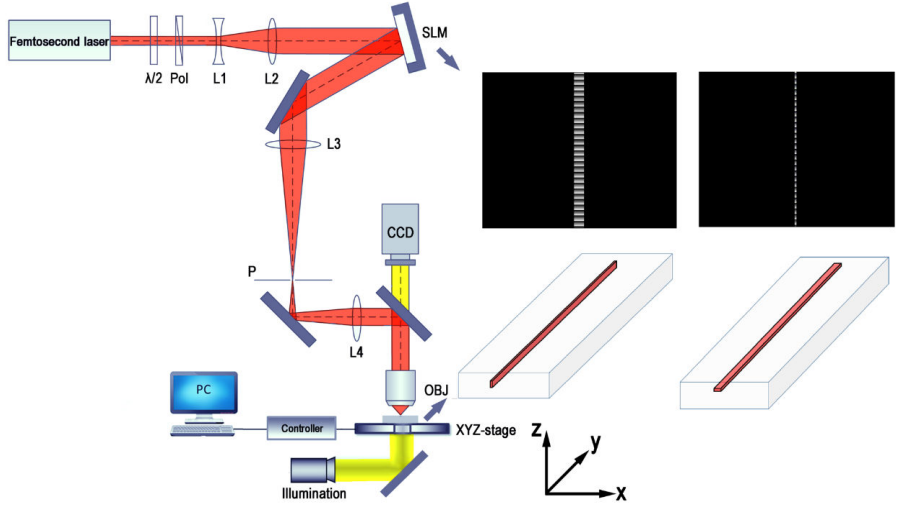

Fig. 1 Schematic illustration of the experimental setup. POL: polarizer. CCD: charge coupled device. OBJ: objective lens. PC: personal computer. L1, L2, L3, and L4 are the lenses of different focal lengths which are described in the main text. Coordinates are indicated in the figure. Inset: example phase masks for writing the vertical and horizontal sides of the cladding, respectively.

In our experiment, commercially available MgO-doped x-cut $LiNbO_3$ crystals of a size of 10 × 5 × 3 $mm^3$ were used as the substrates. Fig. 1 shows the schematic of our experimental setup by which adaptive slit beam shaping was achieved by a spatial light modulator (SLM) [18,19]. The output beam of a Ti:Sapphire laser (Libra-HE, Coherent Inc.) with a maximum pulse energy of 3.5 mJ, an operation wavelength of 800 nm, a pulse width of ~50 fs, and a repetition rate of 1 kHz was used. The pulse energy was controlled by a rotatable half-wave plate and a Glan-Taylor polarizer. The laser beam was first expanded using a concave lens of focal length 10 cm (L1) and a convex lens of focal length 20 cm (L2) before impinging on a reflective phase-only spatial light modulator (SLM, Hamamatsu, X10468-02). To form the slit-shaped beam with the phase-only SLM, a blazed grating with a modulation depth of $2\pi$ rad and a period of 420 μm was used to maximize the diffraction efficiency of the first order. Example phase masks used for writing the horizontal and vertical sides of the cladding are presented in the insets, respectively. The first diffracted order produced by the SLM, after being focused with a lens of focal length 70 cm (L3), was filtered out using a pinhole. The slit-shaped beam were then imaged onto the back aperture of the objective lens (MPLFLN, Olympus) using a lens of focal length 70 cm (L4), and focused into the glass sample to write the horizontal and vertical claddings of buried waveguides. The glass sample was mounted on a computer-controlled XYZ stage with a translation resolution of 1 μm. The fabrication parameters for writing the waveguides of different dimensions are described below in detail when the fabricated structures will be shown.

For waveguides characterization, a fiber laser with a wavelength of 1550 nm (THORLABS S1FC) was used for end-fire coupling into the waveguide. The near–field mode profile at the exit facet is imaged onto an InGaAs camera (Xeva XC-130) with an objective lens of NA = 0.45. To investigate the polarization dependence in the fabricated waveguides, a polarization maintaining fiber (PMF) is used to control the polarization direction of the input beams. Insertion loss measurements are also conducted for all presented waveguides employing a calibrated mid-infrared detector.

## 3. Results and discussions

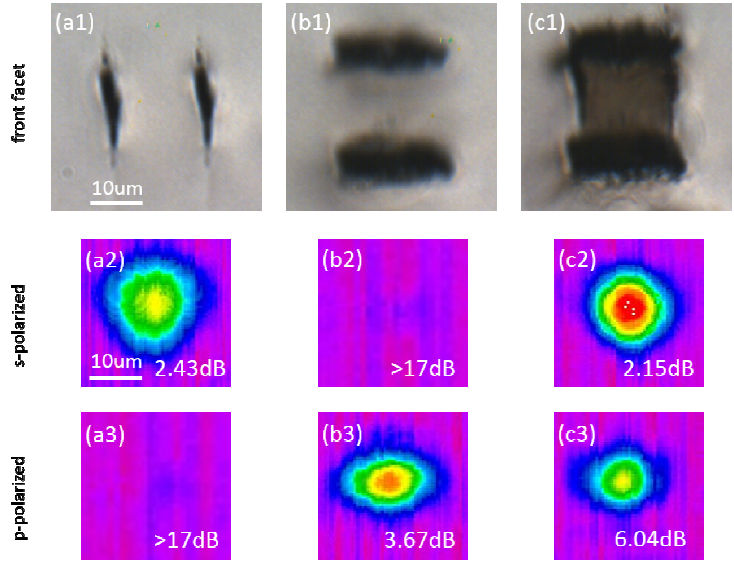

Fig. 2. (first row) Transmitted light microscope images of the different waveguide geometries with a cross-section size of ~13 × 13 μm$^2$ referred to as (a1) vertical double lines, (b1) horizontal double lines, (c1) quadrate four lines. Corresponding mode profiles and insertion loss for (second row) s-, and (third row) p- polarization at λ = 1.55μm, respectively.

The cross-section and the guiding modes of the optical waveguides of different types which are written 60 μm beneath the surface of lithium niobate crystal are compared in Fig. 2. For these waveguides, the distance between the two parallel lines was set to be 13 μm at both vertical and horizontal orientations. Figure 2(a-b) shows the type II waveguides composed of two parallel lines oriented in vertical and horizontal directions, respectively, whereas Fig. 2(c) shows a depressed cladding waveguide which is formed by writing all the four parallel lines as shown in Fig. 2(a) and (b). To write the two vertical tracks of the cladding, the length and width of the slit on the SLM were set to 12 mm and 0.8 mm, respectively. The pulse energy was set to 0.35 μJ. Meanwhile, the two horizontal tracks of the cladding were written by setting the length and width of the slit to 12 mm and 0.12 mm, respectively. The pulse energy was set to 1.80 μJ. The scan speed for the vertical sides of the cladding was set at 0.1 mm/s and scan speed for horizontal sides was set at 0.02 mm/s.

Accordingly, we present the performance of each waveguides for a propagating TE wave in Fig. 2(a2-c2). As shown in Fig. 2(a2-c2), one can see that either the type II waveguide composed of two lines oriented in the vertical direction (Fig. 2(a2)) or the depressed cladding waveguide (Fig. 2(c2)) can guide TE mode, whereas the type II waveguide composed of two lines oriented in the horizontal direction fails in guiding the TE mode wave. In contrast, as we can see in Fig. 2(a3-c3), either the type II waveguide composed of two lines oriented in the horizontal direction (Fig. 2(b3)) or the depressed cladding waveguide (Fig. 2(c3)) can guide TM mode, whereas the type II waveguide composed of two lines oriented in the horizontal direction fails in guiding the TM mode wave. This provides clear evidence that only the depressed cladding waveguide can provide a capability of guiding both TE and TM modes. The size of each guided mode is illustrated as insets in the corresponding figure. It can be seen that for the depressed cladding waveguide, the horizontal and vertical mode-field sizes are ~10 μm × ~10μm for the TE mode, whereas they change to ~9 μm × ~9 μm for the TM mode. The mode-field size can be tuned by changing the distance of the lines as we show below.

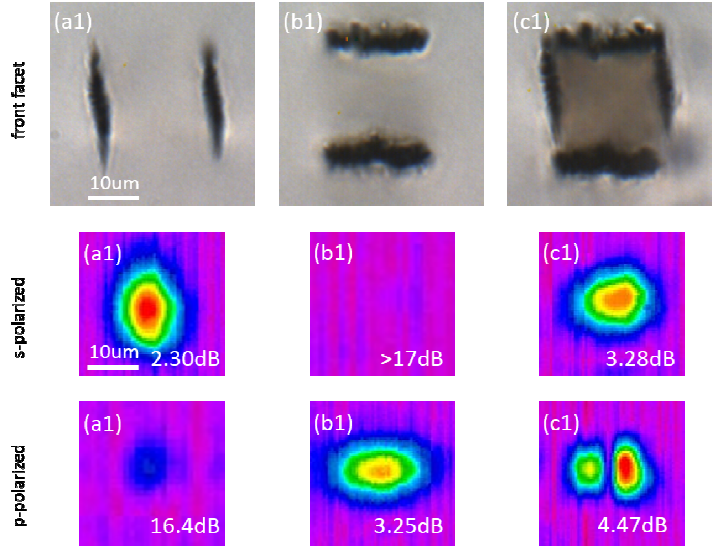

Fig. 3. (first row) Transmitted light microscope images of the different waveguide geometries with a cross-section size of ~18 × 18 μm² referred to as (a1) vertical double lines, (b1) horizontal double lines, (c1) quadrate four lines. Corresponding mode profiles and insertion loss for (second row) s-, and (third row) p- polarization at λ = 1.55μm, respectively.

To experimentally demonstrate the wide range of tuning flexibility in terms of the mode-field size, we fabricated another two parallel lines oriented in vertical and horizontal directions and a depressed cladding waveguides of larger cross sectional sizes of ~18 × 18 μm², as shown in Fig. 3. The lengths of the horizontal and vertical lines can be controlled by varying either the NA of objective lens or the dimensions of the slits. We took the latter option because this requires no mechanical realignment of the optical system. Furthermore, the size of the laser affected zone is also influenced by the laser pulse energy due to the nonlinear self-focusing effect, which provides another control knob for optimizing the dimensions of the the claddings. To write the two vertical tracks of the cladding, the length and width of the slit on the SLM were changed to 12 mm and 0.8 mm, respectively. The pulse energy was changed to 0.38 μJ to ensure a sufficiently high intensity at the focus comparable to that in Fig. 2. Meanwhile, the two horizontal tracks of the cladding were written by setting the length and width of the slit to 12 mm and 0.12 mm, respectively. Accordingly, the pulse energy was changed to 1.91 μJ. The scan speed for the vertical tracks of the cladding was set at 0.1 mm/s and scan speed for horizontal tracks was set at 0.02 mm/s. We note that we choose the same arrangement of the results in Fig. 3 as that in Fig. 2, respectively, except that the waveguides in Fig. 2 are smaller than that of the waveguides in Fig. 3, as clarified above. Therefore, it is clearly seen that the type II waveguide composed of two lines oriented in the vertical direction can only guide the TE mode whereas the type II waveguide composed of two lines oriented in the horizontal direction can only guide the TM mode. Only the depressed cladding waveguides fabricated by simultaneously writing the horizontal and vertical lines can efficiently guide both the TE and TM modes. The larger mode area of the depressed cladding waveguide allows for multi-mode propagation which is illustrated as the insets in the corresponding sub-figures. In contrast to that, the smaller mode area as illustrated in Fig. 2(c) can only allow single-mode propagation.

To estimate the loss in the depressed cladding waveguide, we first measured the power of the laser at 1550 nm wavelength from the output end of the fiber. Afterwards, the laser beam was directly butt coupled into the waveguides without using refractive-index matching liquid. We then measured the output powers from all the depressed cladding waveguides of the different mode-field sizes. From the measured data, the total losses (i.e., including the insertion loss at the entrance of the waveguides and the propagation loss within the waveguides) of the ~13 × 13 $\mu m^2$ depressed cladding waveguide in Fig. 2(c1) are estimated to be 2.15 dB for s-polarized light and 6.04 dB for p-polarized light. In contrast to that, the total losses of the ~18 × 18 $\mu m^2$ cross section in Fig. 3(c1) are estimated to be 3.28dB for s-polarized light and 4.47dB for p-polarized light, which shows less polarization dependence. The origin of the difference will be discussed below. Since it is difficult to independently determine the insertion loss and the propagation loss with high accuracy at the wavelength of 1550 nm with our current equipments, quantitative evaluation of the propagation loss will be carried out with more elaborated methods in the future [20]. Nevertheless, from these total losses measured above which set the up-limits, the propagation losses are reasonable and can already support many applications such as nonlinear photonic circuits and miniaturized lasers.

The waveguiding mechanism of the depressed cladding waveguides proposed by us is slightly different from that of the former type II waveguides. It is widely accepted that the refractive-index distribution in the vertical double-lines waveguide is caused by both a negative refractive-index change in the two laser-written lines and a stress-induced positive refractive-index change between the double lines [21]. The anisotropic stress field induced by the elliptical focus volume related to the vertical double-lines leads to increase of the ordinary refractive index and decrease of the extraordinary refractive index [21]. In our experiment, we have observed exactly the same guiding mode distribution for the perpendicular polarization in the waveguides formed by the horizontal double-lines. However, it can be expected that the stress distribution in the quadrate waveguides will be more isotropic as compared to that in the double-line structures, then the guiding will be mostly contributed by the decrease of the refractive index in the four tracks written by the femtosecond laser (i.e., the depressed cladding), thus allowing for the nearly polarization-independent waveguiding. Quantitative investigations on the stress distribution as a function of the waveguide geometry and laser parameters will be carried out in the future.

## 6. Conclusion

To conclude, we have demonstrated fabrication of depressed cladding waveguides buried in lithium niobate crystal using a transverse femtosecond laser writing scheme. By fully exploring the potential of the slit-beam shaping technique, we show that depressed cladding waveguides of variable cross sectional dimensions can be achieved to enable single-mode propagation of both TE and TM modes in the nonlinear optical crystal. The depressed cladding waveguides inherently prevents the waveguiding areas from modification by the laser irradiation, thereby maintaining all the optical properties of the pristine materials. This is particularly useful for applications including nonlinear photonic circuits and integrated quantum photonic chips.

**Acknowledgments**

We acknowledge Prof. Jiqiao Liu and Mr. Chaolei Yue for their contributions to this research. This work was supported by the National Basic Research Program of China (Grant No. 2014CB921300), National Natural Science Foundation of China (Grant Nos. 61590934, 11134010, and 61327902), and the Youth Innovation Promotion Association of Chinese Academy of Sciences.